\documentclass[a4paper,12pt]{article}

\usepackage{amssymb,amsmath,mathabx,amsthm,amsfonts,mathrsfs,dsfont}
\usepackage{booktabs,colortbl,cases}
\usepackage{relsize}
\usepackage{dsfont}
\usepackage{url,subcaption}
\usepackage[latin1]{inputenc}
\usepackage{eufrak}
\usepackage{epsfig}
\usepackage{color}
\usepackage[noadjust]{cite}
\usepackage{mathrsfs}
\usepackage{tikz}
\usepackage{enumerate}
\usepackage{psfrag}
\usepackage{graphics,adjustbox} 
\usepackage{epsfig} 
\usetikzlibrary{arrows,shapes.geometric,positioning}
\definecolor{my_GREEN}{rgb}{0,0.5,0}
\definecolor{LightCyan}{rgb}{0.88,1,1}
\definecolor{my_Gray}{rgb}{0.85,0.85,0.85}
\title{\LARGE \bf Station Keeping through Beacon-referenced Cyclic Pursuit}
\author{Kevin S. Galloway$^{1}$ and Biswadip Dey$^{2}$
\thanks{*This research was supported in part by the Air Force Office of Scientific Research under AFOSR Grant FA9550-10-1-0250, the ARL/ARO MURI Program Grant W911NF-13-1-0390, and by the Office of Naval Research. Experimental evaluation of the control laws was carried out in the Intelligent Servosystems Laboratory, on a physical test-bed system for synthesis of collective behavior from fundamental building blocks, supported by a FY2012 DURIP Grant from the AFOSR (FA2386-12-1-3002).}
\thanks{$^{1}$Kevin S. Galloway is with the Electrical and Computer Engineering Department, United States Naval Academy, Annapolis, MD 21402 USA. {\tt\small kgallowa@usna.edu}}%
\thanks{$^{2}$Biswadip Dey is with the Department of Electrical and Computer Engineering, Institute for Systems Research, University of Maryland, College Park, MD 20742, USA. {\tt\small biswadip@umd.edu}}%
}

\newtheorem{theorem}{Theorem}[section]

\newtheorem{proposition}[theorem]{Proposition}

\newtheorem{remark}[theorem]{Remark}

\definecolor{myCOLOR1}{RGB}{0,0,255}

\newcommand{\editKG}[1]{{\textcolor{black}{#1}}}

\begin{document}

\maketitle

\begin{abstract}
This paper investigates a modification of cyclic constant bearing (CB) pursuit in a multi-agent system in which each agent pays attention to a neighbor and a beacon. The problem admits shape equilibria with collective circling about the beacon, with the circling radius and angular separation of agents determined by choice of parameters in the feedback law. Stability of circling shape equilibria is shown for a 2-agent system, and the results are demonstrated on a collective of mobile robots tracked by a motion capture system.
\end{abstract}

\section{Introduction}
\label{sec:Intro}
Previous works \cite{Kevin_2011_CDC, Galloway_PRS_13, Marshall_TAC_04, Kim20071426, Smith20051045, Marshall20063, 5160735} have demonstrated pertinence of dyadic pursuit interactions as a building block for collective control. \editKG{In \cite{Kevin_2011_CDC, Galloway_PRS_13}, the particular focus} is on a cyclic pursuit scheme in which each agent employs a constant bearing (CB) pursuit law \cite{Wei_Justh_PSK_09} with regards to exactly one other agent in the collective. In that context, it was demonstrated (for a range of control parameters) that the closed-loop dynamics admit circling relative equilibria (among other special solutions), with the corresponding formation shape determined by the choice of control parameters. However, both the location of the circumcenter (with respect to an inertial frame) and the radius of the circular orbit are determined by initial conditions rather than control parameters, which limits the effectiveness of the control methodology from a design perspective.

In the current work, we introduce a modified version of the CB control law, in which the pursuer is attentive to both a neighboring agent as well as to a beacon (which is assumed to be fixed in the current setting). Though the control law is not designed to stabilize a particular station-keeping range from the beacon, we demonstrate that in an $n$-agent collective where each agent $i$ employs this ``beacon-referenced'' CB control law with respect to agent $i+1$ and a common beacon, circling equilibria exist which are centered on the beacon position and have a radius determined by the control parameters. For the case where $n=2$, we analyze the stability of the associated circling equilibria by linearization of the dynamics, deriving stability conditions in terms of the control parameters. 

While our approach is motivated by the numerous robotic station-keeping applications which require autonomous agents to orbit a specified location while maintaining a fixed formation shape and scale (e.g. search and rescue, environmental sensing, etc.), we also note that this work may provide insights into the mechanisms underlying collective behavior observed in nature. For example, the beacon-referenced cyclic pursuit analyzed in this paper may provide tools for modeling the ``explore-exploit'' behavior evidenced by some animal collectives (e.g. honeybees \cite{Seeley_Behav_Ecol_91}) searching for food sources.  
  
\section{Modeling the interaction}
\label{sec:Model}
%

%
\subsection{Modeling the system}
As presented in \cite{Justh03steeringlaws, Justh_PSK_SCL04}, we model an agent as a unit-mass self-steering particle with twice-differentiable motion path in $\mathds{R}^2$. This allows us to use natural Frenet frame \cite{Nat_Frenet_Bishop} equations to describe the motion for a group of $n$ agents. By letting $\mathbf{r}_i$ denote the position of the $i$-th agent, underlying system dynamics can be expressed as
\begin{equation}
\begin{array}{rcl}
\dot{\mathbf{r}}_i &=& \nu_i \mathbf{x}_i \\
\dot{\mathbf{x}}_i &=& \nu_i u_i \mathbf{y}_i \\
\dot{\mathbf{y}}_i &=& - \nu_i u_i \mathbf{x}_i, \quad i = 1,2,\ldots,n.
\end{array} \label{Explicit_MODEL}
\end{equation}
Here $\mathbf{x}_i$ is the normalized velocity and $\mathbf{y}_i$ represents orthogonally rotated $\mathbf{x}_i$ in the counter-clockwise direction. Moreover, $\nu_i$ is the speed, and $u_i$ is the natural curvature viewed as a steering control. Alternatively, by packing $\mathbf{r}_i, \mathbf{x}_i, \mathbf{y}_i$ inside a matrix
\begin{equation}
g_i 
\triangleq 
\left[ \begin{array}{ccc}
\mathbf{x}_i & \mathbf{y}_i & \mathbf{r}_i \\
0 & 0 & 1
\end{array} \right]
\in SE(2),
\label{GrP_Defn_SE2}
\end{equation}
the natural Frenet frame equations \eqref{Explicit_MODEL} can be expressed as a left invariant dynamics on $SE(2)$. From this perspective, the dynamics of an agent can be expressed as
\begin{equation}
\dot{g}_i = g_i\xi_i = \nu_i g_i (X_0 + u_iX_2),
\label{Implicit_MODEL}
\end{equation}
where $X_0$ and $X_2$ represent standard basis elements of $\mathfrak{se}(2)$.

As discussed earlier, practical applications often require the collective of agents to maneuver with respect to some particular desired location. Therefore, in this work we introduce a beacon at location $\mathbf{r}_b \in \mathds{R}^2$, along with a fixed frame $[\mathbf{x}_b \; \mathbf{y}_b]$ attached to it (which is assumed to be the inertial reference frame, without loss of generality), and define $g_b \in SE(2)$ as in \eqref{GrP_Defn_SE2} to pack $\mathbf{r}_b$, $\mathbf{x}_b$ and $\mathbf{y}_b$ inside a single matrix.
\subsection{Directed graph and commutativity constraints}
Now we introduce the notion of an \textit{attention graph} \cite{Galloway_PRS_13} to describe which agent(s) a particular agent is paying attention to. By letting $\mathcal{N} = \{1,2,\ldots,n,b\}$ denote the node-set for the problem of our concern, the corresponding arc/edge-set ($\mathcal{A}$) is defined as
\begin{equation}
\mathcal{A}
=
\big\{
(i,i+1), (i,b) 
\big|
i = 1,2,\ldots ,n
\big\},
\end{equation}
where addition in the index variables should be interpreted modulo $n$ throughout this paper. Clearly, this attention graph $\mathcal{G} = (\mathcal{N}, \mathcal{A})$ is weakly connected and devoid of any self loop. 

Now, we formulate a reduction to the shape space, and introduce the following set of variables (possibly redundant) along arcs of the attention graph $\mathcal{G}$:
\begin{align}
\tilde{g}_{i,i+1} &= g_{i+1}^{-1} g_i
\label{GrP_Defn_Shape_1}
\\
\textrm{and,} \qquad
\tilde{g}_{ib} &= g_b^{-1} g_i.
\label{GrP_Defn_Shape_2}
\end{align}
It follows from the definition that $\tilde{g}_{i,i+1}$, $i = 1,2,\ldots,n$ are subject to the cycle closure constraint
\begin{equation}
\prod\limits_{i=1}^n \tilde{g}_{i,i+1}
=
\tilde{g}_{n1}\tilde{g}_{n-1,n} \cdots \tilde{g}_{23} \tilde{g}_{12}
=
\mathds{I}_3,
\label{constraint_CYCLE}
\end{equation}
where $\mathds{I}_3$ is the $3\times 3$ identity matrix. (See also \cite{Galloway_PRS_13}.)  Moreover, by exploiting the commutative property, we have
\begin{equation}
\tilde{g}_{i,i+1} = \tilde{g}_{i+1,b}^{-1} \tilde{g}_{ib}, \qquad i = 1,2,\ldots,n-1.
\label{constraint_GROUP}
\end{equation}
This set of equations \eqref{constraint_GROUP} poses consistency conditions on the space of shape variables. We should note here that together \eqref{constraint_CYCLE} and \eqref{constraint_GROUP} ensure $\tilde{g}_{n1} = \tilde{g}_{1b}^{-1} \tilde{g}_{nb}$.

Also, from \eqref{Implicit_MODEL}, \eqref{GrP_Defn_Shape_1} and \eqref{GrP_Defn_Shape_2}, it follows that the shape dynamics can be expressed as 
\begin{equation}
\begin{aligned}
\dot{\tilde{g}}_{ib} &= \tilde{g}_{ib} \xi_i,
\\
\textrm{and} \quad \dot{\tilde{g}}_{i,i+1} &= \tilde{g}_{i,i+1} \tilde{\xi}_{i,i+1},
\end{aligned}
\label{Shape_dynamics_Implicit}
\end{equation} 
where $\tilde{\xi}_{i,i+1} = \xi_i - \tilde{g}^{-1}_{i,i+1} \xi_{i+1} \tilde{g}_{i,i+1}$. It is a straightforward exercise to show that the constraints \eqref{constraint_CYCLE}-\eqref{constraint_GROUP} will be satisfied for all future time if they are satisfied initially.
%
%
%
%

%
\subsection{Scalar shape variables}
\begin{figure}[h]
\begin{center}
  \includegraphics[width=0.48\textwidth]{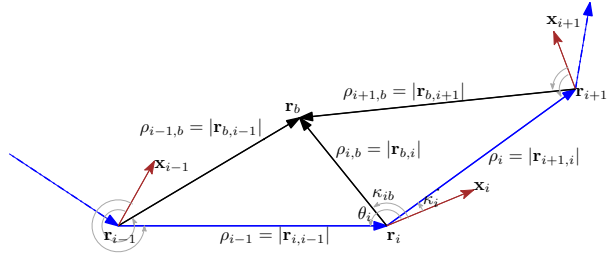}
  \caption{Illustration of the scalar shape variables ($\rho_i$, $\rho_{ib}$, $\theta_i$, $\kappa_i$ and $\kappa_{ib}$) used in our analysis of an $n$-agent system with a fixed beacon at location $\mathbf{r}_b$.} 
  \label{Scalar_Shapes}
\end{center}
\end{figure}

Now, following the approach of earlier works \cite{Galloway_PRS_13}, we introduce scalar shape variables (a polar parametrization) to describe the state of an agent relative to the beacon and other agents. By letting \begin{equation}
R(\vartheta)
=
\left[
\begin{array}{cc}
\cos\vartheta & -\sin\vartheta \\ \sin\vartheta & \cos\vartheta
\end{array}\right]
\in SO(2)
\end{equation}
denote a counter-clockwise planar rotation through an angle $\vartheta$, we define a set of scalar shape variables $\rho_i$, $\rho_{ib}$, $\kappa_i$, $\theta_i$, $\kappa_{ib}$ and $\psi_i$ as
\begin{align}
& \rho_i = |\mathbf{r}_{i+1,i}|,
\qquad 
&& \rho_{ib} = |\mathbf{r}_{b,i}|,
\nonumber \\
& R(\kappa_i)\mathbf{x}_i = \frac{\mathbf{r}_{i+1,i}}{|\mathbf{r}_{i+1,i}|}, 
\qquad
&& R(\theta_i)\mathbf{x}_i = -\frac{\mathbf{r}_{i,i-1}}{|\mathbf{r}_{i,i-1}|}
\label{Scalar_Shape_DEFN} \\
& R(\kappa_{ib})\mathbf{x}_i = \frac{\mathbf{r}_{b,i}}{|\mathbf{r}_{b,i}|},
&& R(\pi - \psi_i)\mathbf{x}_b = \frac{\mathbf{r}_{b,i}}{|\mathbf{r}_{b,i}|},
\nonumber
\end{align}
where $i=1,2,\ldots,n$ and $\mathbf{r}_{i,j} = \mathbf{r}_{i} - \mathbf{r}_{j}$ represents the position of $i$-th agent relative to the $j$-th agent (see Fig~\ref{Scalar_Shapes}). 

Using these recently introduced scalar shape variables, the consistency conditions \eqref{constraint_GROUP} can be expressed as 
\begin{align}
&
R(\psi_i - \psi_{i+1}) 
=
R(\pi + \kappa_i - \theta_{i+1} + \kappa_{i+1,b} - \kappa_{ib})
\label{constraint_1_SHAPE}
\\
&
\rho_i \mathds{I}_2 = \rho_{ib}R(\kappa_{ib} - \kappa_i) + \rho_{i+1,b}R(\kappa_{i+1,b} - \theta_{i+1})
\label{constraint_2_SHAPE_final}
\end{align}
for $i = 1,2,\ldots,n-1$. In a similar way, the cycle closure constraint \eqref{constraint_CYCLE} can be expressed as
\begin{align}
&
R\big( \sum\limits_{i = 1}^{n} (\pi + \kappa_i - \theta_{i+1}) \big)
=
\mathds{I}_2
\label{Cycle_Closure_1}
\\
&
\sum\limits_{i=1}^n \rho_i R\big( \sum\limits_{j = 1}^{i} (\pi + \kappa_j - \theta_{j+1}) \big)
=
0.
\label{Cycle_Closure_2}
\end{align}
In actuality, certain simplifications allow us to consider only a subset of these variables and constraints. First, it is clear that for a particular choice of $\psi_1$, \eqref{constraint_1_SHAPE} provides an explicit representation for the scalar shape variables $\psi_i$, $i = 2,3,\ldots,n$ in terms of other shape variables, and therefore we can disregard those variables as well as constraint \eqref{constraint_1_SHAPE} from the following analysis. Furthermore, the choice of beacon frame $[\mathbf{x}_b \; \mathbf{y}_b]$ is arbitrary and the proposed feedback law (see Section~III) is invariant to any rotation of the beacon frame, and therefore there exists an $\mathcal{S}^1$ symmetry enabling us to exclude $\psi_1$ from further analysis. Finally, one can show that \eqref{constraint_2_SHAPE_final} together with \eqref{Cycle_Closure_1} and \eqref{Cycle_Closure_2} can be expressed more concisely as simply \eqref{constraint_2_SHAPE_final} (holding for $i=1,2,\ldots,n$) with \eqref{Cycle_Closure_1}. Therefore, the shape space can be parametrized by the scalar variables $\kappa_i, \kappa_{ib}, \theta_i, \rho_i, \rho_{ib}$, for $i=1,2,\ldots,n$, subject only to the positivity constraints on $\rho_{i}$ and $\rho_{ib}$ (necessary for the well-posedness of a feedback law introduced in the next section) and the constraints \eqref{constraint_2_SHAPE_final} - \eqref{Cycle_Closure_1} for $i=1,2,\ldots,n$, i.e.
\begin{align}
&
R\big( \sum\limits_{i = 1}^{n} (\pi + \kappa_i - \theta_{i+1}) \big)
=
\mathds{I}_2
\label{FINAL_Constraint_1}
\\
&
\rho_i \mathds{I}_2 = \rho_{ib}R(\kappa_{ib} - \kappa_i) + \rho_{i+1,b}R(\kappa_{i+1,b} - \theta_{i+1}).
\label{FINAL_Constraint_2}
\end{align}

Next we focus on representing the shape dynamics \eqref{Shape_dynamics_Implicit} in terms of scalar shape variables. By straightforward calculations, one can show that the shape dynamics are given in terms of the scalar variables by
\begin{align}
\dot{\rho}_i 
&= 
- \nu_i\cos\kappa_i - \nu_{i+1}\cos\theta_{i+1}
\nonumber \\
\dot{\kappa}_i 
&= 
- \nu_i u_i + \frac{1}{\rho_i}\big[ \nu_i\sin\kappa_i + \nu_{i+1}\sin\theta_{i+1}\big]
\nonumber \\
\dot{\theta}_i 
&= 
- \nu_i u_i + \frac{1}{\rho_{i-1}}\big[ \nu_{i-1} \sin\kappa_{i-1} + \nu_i \sin\theta_{i}\big]
\label{shape_dynamics_EXPLICIT} \\
\dot{\rho}_{ib} 
&= 
- \nu_{i}\cos\kappa_{ib}
\nonumber \\
\dot{\kappa}_{ib} 
&= 
- \nu_i u_i + \frac{\nu_i}{\rho_{ib}} \sin\kappa_{ib}, \; i=1,2,\ldots,n,
\nonumber
\end{align}
subject to the cycle closure constraint \eqref{FINAL_Constraint_1} and consistency conditions \eqref{FINAL_Constraint_2}.
\section{\editKG{A beacon-referenced CB pursuit law}}
\label{sec:CL_Dyna}
%

%
In this section we introduce a modified version of the CB pursuit law from \cite{Wei_Justh_PSK_09} that introduces an additional term referenced to the beacon bearing. We construct this feedback law as a convex combination of two fundamental building blocks, expressed as
\begin{equation}
u_i = (1 - \lambda)u_{CB}^i + \lambda u_B^i, \qquad \lambda \in [0,1]
\label{u_i_top_level}
\end{equation}
where $u_{CB}^i$ is given by the original CB pursuit law \cite{Wei_Justh_PSK_09} referenced to agent $i+1$, and $u_B^i$ represents the deviation from a desired bearing angle to the beacon. More specifically, we choose
\begin{align}
u_{CB}^i
&=
- \mu_i \left( R(\alpha_i)\mathbf{y}_i \cdot \frac{\mathbf{r}_{i,i+1}}{|\mathbf{r}_{i,i+1}|} \right)
- \frac{1}{\nu_i |\mathbf{r}_{i,i+1}|} \left( \frac{\mathbf{r}_{i,i+1}}{|\mathbf{r}_{i,i+1}|} \cdot R(\pi/2) \dot{\mathbf{r}}_{i,i+1} \right),
\label{u_CB_i}
\end{align}
with $\mu_i > 0$ being a control gain and the angle $\alpha_i \in S^1$ representing the desired offset between the $i$-th agent's heading and the current location of the $(i+1)$-th agent. For the beacon tracking component, we let
\begin{equation}
u_B^i
=
- \mu_i^b \left( R(\alpha_{ib})\mathbf{y}_i \cdot \frac{\mathbf{r}_{i,b}}{|\mathbf{r}_{i,b}|} \right),
\label{u_Beacon_i}
\end{equation}
where $\mu_i^b > 0$ is the corresponding control gain and the angle $\alpha_{ib} \in S^1$ is the desired offset between the current heading of the $i$-th agent and the bearing to the beacon location. Note that the neighbor tracking goal may conflict with the beacon referencing goal, and the parameter $\lambda$ maintains a balance between the beacon's influence and the influence of the neighboring agent $i+1$. In particular, for $\lambda = 0$, $u_i$ is simply the original CB pursuit law with no reference to the beacon.

In terms of scalar shape variables, the feedback law $u_i$ can be expressed as
\begin{align}
u_i
&=
\lambda \mu_i^b\sin(\kappa_{ib} - \alpha_{ib})
+
(1 - \lambda)\mu_i\sin(\kappa_i - \alpha_i) 
+ \frac{1 - \lambda}{\rho_i}\left(\sin\kappa_i + \frac{\nu_{i+1}}{\nu_i}\sin\theta_{i+1}\right).
\label{u_i_shape}
\end{align}

\begin{remark}
As noted in \cite{Wei_Justh_PSK_09}, the last component of this feedback law \eqref{u_i_shape} can be interpreted as the angular speed at which the baseline between agent-$i$ and agent-$i+1$ is rotating around the $i$-th agent. Therefore it is plausible to evaluate the steering command $u_i$ without explicit measurement of distance between the agents, although it will require an appropriate sensing mechanism (mimicking the principle of compound eyes in visual insects).
\end{remark}

Before going into detailed analysis of relative equilibria, we introduce the following simplifying assumptions:
\begin{itemize}
\item[(A1)] The speed of the agents are equal and constant. Hence, without loss of generality, we can assume $\nu_i = 1$ for every $i \in \{1,2,\ldots,n\}$.
\item[(A2)] The controller gains ($\mu_i$ and $\mu_i^b$) are equal and common for all agents, i.e. $\mu_i = \mu_i^b = \mu $ for every $i \in \{1,2,\ldots,n\}$.
\item[(A3)] The bearing angles ($\alpha_{ib}$) with respect to the beacon are common for all agents, i.e. $\alpha_{ib} = \alpha_0 $ for every $i \in \{1,2,\ldots,n\}$.
\end{itemize}
Under assumptions (A1)-(A3), the closed loop shape dynamics \editKG{(\eqref{shape_dynamics_EXPLICIT} with \eqref{u_i_shape})} can be expressed as
\begin{align}
\dot{\rho}_i 
&= 
- \big( \cos\kappa_i + \cos\theta_{i+1} \big)
\nonumber \\
\dot{\kappa}_i 
&= 
- \mu \big[ (1 - \lambda) \sin(\kappa_i - \alpha_i) + \lambda \sin(\kappa_{ib} - \alpha_0) \big] 
+ \frac{\lambda}{\rho_i} \big[ \sin\kappa_i + \sin\theta_{i+1}\big]
\nonumber \\
\dot{\theta}_i 
&= 
\dot{\kappa}_i 
- \frac{1}{\rho_i}\big[ \sin\kappa_i + \sin\theta_{i+1}\big]
+ \frac{1}{\rho_{i-1}}\big[ \sin\kappa_{i-1} + \sin\theta_{i}\big]
\label{CL_dynamics_n_simplified} \\
\dot{\rho}_{ib} 
&= 
- \cos\kappa_{ib}
\nonumber \\
\dot{\kappa}_{ib} 
&= 
\dot{\kappa}_i 
- \frac{1}{\rho_i}\big[ \sin\kappa_i + \sin\theta_{i+1}\big]
+ \frac{1}{\rho_{ib}} \sin\kappa_{ib}
\nonumber
\end{align}
for $i = 1,2,\ldots,n$, subject to the constraints \eqref{FINAL_Constraint_1}-\eqref{FINAL_Constraint_2}.
\section{Relative equilibria}
\label{sec:Rel_EQ_exist}
%

%
In this section we analyze the closed loop shape dynamics \eqref{CL_dynamics_n_simplified} to determine existence conditions and characterization of their equilibria (i.e. relative equilibria for the full dynamics \eqref{Explicit_MODEL} with \eqref{u_i_top_level}). At the extreme value of $\lambda = 0$, the shape dynamics simplify to the cyclic CB pursuit dynamics previously analyzed in \cite{Galloway_PRS_13}, while at the other extreme, inter-agent interaction is completely lost whenever $\lambda = 1$. Therefore we restrict $\lambda$ to lie in the open interval $(0,1)$ for the rest of our analysis.

From the form of $\dot{\rho}_{ib}$ and $\dot{\rho}_i$ in \eqref{CL_dynamics_n_simplified}, we can obtain necessary conditions at equilibrium given by
\begin{equation}
\kappa_{ib} = \pm \frac{\pi}{2},
\quad \textrm{and,} \quad 
\theta_{i+1} = \pi \pm \kappa_i,
\label{eq_kappa_i_ib}
\end{equation}
for $i = 1,2,\ldots,n$. Similarly, by setting the dynamics of $\theta_i$ and $\kappa_{ib}$ equal to zero, we obtain (for $i = 1,2,\ldots,n$)
\begin{align}
\frac{1}{\rho_i}(\sin\kappa_i + \sin\theta_{i+1})
&=
\frac{1}{\rho_{i-1}}(\sin\kappa_{i-1} + \sin\theta_{i}),
\label{2nd_One}
\\
\frac{1}{\rho_i}(\sin\kappa_i + \sin\theta_{i+1})
&=
\frac{1}{\rho_{ib}} \sin\kappa_{ib}.
\label{3rd_One}
\end{align}
A straightforward calculation reveals that if $\kappa_{ib} = \pm \pi/2$ with $\kappa_i = \pi + \theta_{i+1}$, then \eqref{3rd_One} results in a contradiction since its left hand side vanishes to zero, contrary to a non-zero ($\pm 1/\rho_{ib}$) right hand side. Thus, at a relative equilibrium, we have
\begin{equation}
\theta_{i+1} = \pi - \kappa_i,
\quad 
i = 1,2,\ldots,n.
\label{theta_soln}
\end{equation}
By introducing a new variable $\gamma_i$ defined as
\begin{equation}
\gamma_i \triangleq \frac{1}{\rho_i}\big(\sin\kappa_i + \sin\theta_{i+1}\big) = \frac{2}{\rho_i} \sin\kappa_i,
\end{equation}
we obtain the following condition
\begin{equation}
\gamma_i = \gamma_{i-1}
\qquad 
i = 1,2,\ldots,n 
\label{eq_gamma_constant}
\end{equation}
from \eqref{2nd_One}. This condition, along with \eqref{3rd_One}, gives rise to
\begin{equation}
\frac{\sin\kappa_{ib}}{\rho_{ib}} = \frac{\sin\kappa_{i-1,b}}{\rho_{i-1,b}}
\qquad 
i = 1,2,\ldots,n,
\end{equation}
which in turn yields the equilibrium values of $\kappa_{ib}$ as
\begin{equation}
\kappa_{ib}
=
\left\{
\begin{array}{ll}
 \pi/2 & \quad \forall i \in \{1,2,\ldots,n\}, \qquad \textrm{or} \\
-\pi/2 & \quad \forall i \in \{1,2,\ldots,n\}.
\end{array}
\right.
\label{kappa_ib_soln}
\end{equation}

Now setting the dynamics of $\kappa_i$ to zero, we obtain
\begin{align}
&\mu \Big[(1 - \lambda)\sin(\kappa_i - \alpha_i) + \lambda \sin(\kappa_{ib} - \alpha_0) \Big]
\nonumber \\
& \qquad 
=  \frac{\lambda}{\rho_i}(\sin\kappa_i + \sin\theta_{i+1})
= \gamma_i
\label{1st_One}
\end{align}
for $i = 1,2,\ldots,n$. As $\kappa_{ib} = \pm\pi/2$ and $\theta_{i+1} = \pi - \kappa_i$ at a relative equilibrium, \eqref{1st_One} yields an equilibrium value for $\rho_i$ given by
\begin{equation}
\rho_i
=
\frac
{2 \sin\kappa_{ib} \sin\kappa_i}
{\mu \left( \frac{1}{\lambda} - 1 \right) \sin\kappa_{ib} \sin(\kappa_i - \alpha_i) + \mu \cos \alpha_0 }.
\label{eq_rho_i}
\end{equation}
Similarly, by \eqref{3rd_One} an equilibrium value of $\rho_{ib}$ can be expressed as
\begin{equation}
\rho_{ib}
=
\frac{1}{\mu \left( \frac{1}{\lambda} - 1 \right) \sin\kappa_{ib} \sin(\kappa_i - \alpha_i) + \mu \cos \alpha_0}
=
\frac{\sin\kappa_{ib}}{\gamma_i}.
\label{eq_rho_ib_equidistant}
\end{equation}
As we have shown earlier in \eqref{eq_gamma_constant} that $\gamma_i = \gamma_{i-1}$ for every $i \in \{1,2,\ldots,n\}$, it follows from \eqref{eq_rho_ib_equidistant} that the agents will be equi-distant from the beacon at any relative equilibrium. Hence, \textbf{all relative equilibria for the system must be circling equilibria}. Moreover, as both $\rho_{i}$ and $\rho_{ib}$ are required to be positive, we have \textit{necessary} conditions for existence of circling equilibrium, given by
\begin{align}
&
\lambda \cos \alpha_0 + (1 - \lambda) \sin \kappa_{ib} \sin(\kappa_i - \alpha_i) > 0
\label{positive_1}
\\
\textrm{and,} \qquad &
\sin \kappa_{ib} \sin\kappa_i > 0.
\label{positive_2}
\end{align}

\subsection{Evaluating solutions for $\kappa_i$}

It is easy to check that at any (circling) equilibrium of the closed loop shape dynamics,  \eqref{eq_gamma_constant} with \eqref{1st_One} implies that
\begin{equation}
\sin(\kappa_{i+1} - \alpha_{i+1})
=
\sin(\kappa_i - \alpha_i),
\label{angle_condition_n}
\end{equation}
for which we have two possible solutions given by
\begin{subnumcases}{\kappa_{i+1} - \alpha_{i+1} = }
\kappa_i - \alpha_i
\label{Solution_normal}
\\
\pi - (\kappa_i - \alpha_i)
\label{Solution_abnormal}
\end{subnumcases}
where $i \in \{1,2,\ldots,n\}$. Equilibrium values for $\kappa_i$ can therefore be derived from \eqref{Solution_normal}-\eqref{Solution_abnormal}, along with shape variable constraints \eqref{FINAL_Constraint_1}-\eqref{FINAL_Constraint_2} and positivity conditions \eqref{positive_1}-\eqref{positive_2}. 

If we let $\alpha^*$ be the angle satisfying $\kappa_1 - \alpha_1 = \alpha^*$ \editKG{at equilibrium}, then by \eqref{Solution_normal}-\eqref{Solution_abnormal} $\kappa_2 - \alpha_2$ must assume either $\alpha^*$ or $\pi - \alpha^*$. This aspect of binary possibilities holds true for every agent and therefore results in the branching depicted in Fig~\ref{fig:tree_1}. This figure provides a graphical illustration of all possible solutions for \eqref{angle_condition_n}, with each branch representing a candidate solution for $\kappa_i$. By considering a particular branch of the tree with $M$ (where $M \in \{1,2,\cdots,n\}$) copies of $\alpha^*$ and $(n-M)$ copies of $\pi - \alpha^*$, we have 
\begin{equation}
\sum\limits_{i=1}^n \kappa_i
= 
(n-M)\pi + (2M-n)\alpha^* + \sum\limits_{i=1}^n \alpha_i.
\label{sum_kappa_alpha_diff}
\end{equation}
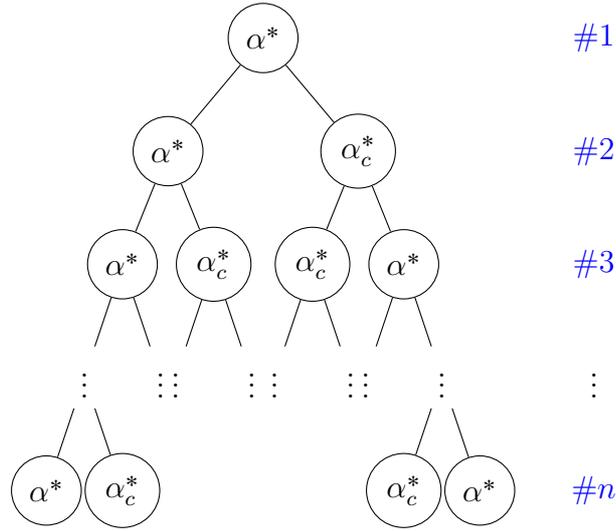
\begin{figure}[t!]
\centering
\begin{tikzpicture}[
	level 1/.style={sibling distance=25mm},
	level 2/.style={sibling distance=12mm},
	level 3/.style={sibling distance=10mm},
	level 4/.style={sibling distance=10mm}		
	]
\node [circle,draw] (z){$\alpha^*$}
  child {node [circle,draw] (a) {$\alpha^*$}
    child {node [circle,draw] (b) {$\alpha^*$}
      child {node {$\vdots$}
        child {node [circle,draw] (d) {$\alpha^*$}}
        child {node [circle,draw] (e) {$\alpha_c^*$}}
      } 
      child {node {$\vdots$}}
    }
    child {node [circle,draw] (g) {$\alpha_c^*$}
      child {node {$\vdots$}}
      child {node {$\vdots$}}
    }
  }
  child {node [circle,draw] (j) {$\alpha_c^*$}
    child {node [circle,draw] (k) {$\alpha_c^*$}
      child {node {$\vdots$}}
      child {node {$\vdots$}}
    }
  child {node [circle,draw] (l) {$\alpha^*$}
    child {node {$\vdots$}}
    child {node (c){$\vdots$}
      child {node [circle,draw] (o) {$\alpha_c^*$}}
      child {node [circle,draw] (p) {$\alpha^*$}
          child [grow=right] {node (q) {$\color{blue} \# n$} edge from parent[draw=none]
            child [grow=up] {node (r) {$\vdots$} edge from parent[draw=none]
              child [grow=up] {node (s) {$\color{blue} \# 3$} edge from parent[draw=none]
                child [grow=up] {node (t) {$\color{blue} \# 2$} edge from parent[draw=none]
                  child [grow=up] {node (u) {$\color{blue} \# 1$} edge from parent[draw=none]}
                }
              }
          }
        }
      }
    }
  }
};
\end{tikzpicture}
\caption{Graphic representation of \textit{all possible} solutions for $\sin(\kappa_{i+1}-\alpha_{i+1}) = \sin(\kappa_i-\alpha_i)$, $i = 1,2,\ldots,n$ wherein $\alpha_c^* = \pi - \alpha^*$. It is important to note here that a particular branch of this tree might not yield a plausible value of $\kappa_i$ (due to closure and positivity constraints). The leftmost branch in this tree represents the relative equilibrium with equal $(\kappa_i-\alpha_i)$ for every agent.} 
\label{fig:tree_1}
\end{figure}
%

Now we focus on the shape variable constraints \eqref{FINAL_Constraint_1}-\eqref{FINAL_Constraint_2} to obtain solutions for $\alpha^*$. It is easy to check that the consistency conditions \eqref{FINAL_Constraint_2} hold true at any relative equilibrium of the closed loop dynamics. Additionally, by exploiting the relationship between equilibrium values of $\kappa_i$ and $\theta_i$, the cycle closure constraint \eqref{FINAL_Constraint_1} can be expressed as
\begin{equation}
\sum_{i=1}^{n} \kappa_i
=
m\pi, \qquad m \in \mathds{Z},
\label{Closure_Bye_Product_1}
\end{equation}
where $\mathds{Z}$ is the set of integers. By substituting \eqref{Closure_Bye_Product_1} into \eqref{sum_kappa_alpha_diff}, we have
\begin{equation}
(2M-n)\alpha^*
= 
(m+M-n)\pi - \sum_{i=1}^n\alpha_i .
\label{cond_on_alpha_star}
\end{equation}

We summarize the preceding discussion in the following proposition.

\begin{proposition}
\label{prop:existenceProp}
Consider an $n$-agent cyclic CB pursuit system with beacon, whose shape dynamics is governed by \eqref{CL_dynamics_n_simplified} \editKG{parametrized by $\mu, \lambda$, and the CB parameters $\left\{\alpha_0, \alpha_1, \alpha_2, \ldots, \alpha_n \right\}$}. The following statements are true.
\begin{itemize}
\item[(a)] The only possible relative equilibria are circling equilibria.
\item[(b)] Whenever $\sin(\sum \alpha_i) \neq 0$, a circling equilibrium exists \textit{if and only if} there exists $m \in \mathds{Z}$ and $\sigma = (\sigma_1,\sigma_2,\ldots,\sigma_n) \in \{-1,1\}^n$ such that 
\begin{itemize}
\item[(i)] \editKG{the cardinality $M$ of the subset $\{\sigma_i | \sigma_i = 1, i = 1,2,\ldots,n\}$ satisfies}
\begin{equation}
2M-n \neq 0,
\label{Prop_4_cond_on_M}
\end{equation}
and
\item[(ii)]
\begin{equation}
\begin{aligned}
& \lambda \cos \alpha_0 + (1 - \lambda) \sin \alpha^* > 0,
\\
& \sin \big( \alpha^* + \sigma_i\alpha_i \big) > 0, \qquad i = 1,2,\ldots,n
\end{aligned}
\label{Proposition_4}
\end{equation}
where $\alpha^*$ is given by
\begin{equation}
\alpha^* = \left( \frac{m+M-n}{2M-n} \right)\pi - \sum_{i=1}^{n} \left( \frac{\alpha_i}{2M-n} \right).
\label{alpha_star_soln}
\end{equation}
\end{itemize}
At equilibrium, we have \editKG{either $\kappa_{ib} = \pi/2, \; i=1,2,\ldots, n$ or $\kappa_{ib} = -\pi/2, \; i=1,2,\ldots, n$} and the equilibrium values of $\kappa_i$, $\rho_{ib}$, and $\rho_i$ given by 
\begin{align*}
\kappa_i
&=
\editKG{\frac{\pi (1-\sigma_i)}{2} + (\sigma_i \alpha^* + \alpha_i) = \;}
\biggl\{
\begin{array}{ll}
\alpha^* + \alpha_i, &\qquad \text{if } \sigma_i = +1 \\
\pi - \alpha^* + \alpha_i, &\qquad \text{if } \sigma_i = -1
\end{array}\\
\rho_{ib}
&=
\frac{1}{\mu \lambda \cos \alpha_0 + \mu \left( 1 - \lambda \right) \sin\kappa_{ib} \sin\alpha^*}\\
\rho_{i}
&=
2\rho_{ib} \sin \kappa_{ib} \sin \kappa_i .
\end{align*}
\end{itemize}
\end{proposition}
\vspace{.3cm}

\begin{proof}
The first statement of the proposition directly follows from \eqref{eq_rho_ib_equidistant}.

\editKG{The preceding discussion has demonstrated that if a circling equilibrium exists, then $\kappa_{ib}$ and $\kappa_i$ must satisfy \eqref{kappa_ib_soln},\eqref{positive_1}-\eqref{positive_2}, \eqref{Solution_normal}-\eqref{Solution_abnormal}, and \eqref{Closure_Bye_Product_1}, and the equilibrium values for $\theta_i$, $\rho_i$\, and $\rho_{ib}$ can be expressed in terms of $\kappa_{ib}$ and $\kappa_i$ by \eqref{theta_soln}, \eqref{eq_rho_i}, and \eqref{eq_rho_ib_equidistant}. Further analysis of \eqref{positive_1}-\eqref{positive_2} demonstrated that if we let $\alpha^*$ denotes the angle difference $(\kappa_1 - \alpha_1)$, then for each $i$, we have one of two possibilities - either $\kappa_i = \alpha^* + \alpha_i$ or $\kappa_i = \pi - \alpha^* + \alpha_i$, which we represent by the binary tree in Fig~\ref{fig:tree_1}.} 


We now consider a particular branch of the binary tree (Fig~\ref{fig:tree_1}), for which $\kappa_i = \alpha^* + \alpha_i$ for exactly $M$ agents ($1 \leq M \leq n$) and $\kappa_i = \pi - \alpha^* + \alpha_i$ for the remaining $n-M$ agents. Clearly, for the first set of $M$ agents we have $\sin \kappa_i = \sin (\alpha^* + \alpha_i)$, while the remaining agents will have $\sin \kappa_i = \sin (\alpha^* - \alpha_i)$, and therefore there exists some $\sigma = (\sigma_1,\sigma_2,\ldots,\sigma_n) \in \{-1,1\}^n$ such that $\sin\kappa_i = \sin (\alpha^* + \sigma_i \alpha_i)$ for every $i=1,2,\ldots,n$, and the cardinality of the set $\{\sigma_i | \sigma_i = 1, i = 1,2,\ldots,n\}$ is $M$. \editKG{This implies that \eqref{cond_on_alpha_star} will hold (as demonstrated in the previous discussion), and thus} for $\kappa_{ib} = \pi/2$, it is clear that the positivity conditions \eqref{positive_1}-\eqref{positive_2} can be expressed as \eqref{Proposition_4}, with \eqref{alpha_star_soln} following from \eqref{cond_on_alpha_star} as long as $2M - n \neq 0$. 

It remains to be shown that \eqref{Proposition_4} also encompasses the case $\kappa_{ib} = -\pi/2$, so that \eqref{Proposition_4} is equivalent to \eqref{positive_1}-\eqref{positive_2}. For $\kappa_{ib} = -\pi/2$, the positivity conditions \eqref{positive_1}-\eqref{positive_2} simplify to
\begin{equation}
\begin{aligned}
& 
\lambda \cos \alpha_0 - (1 - \lambda) \sin \alpha^* > 0
\\
&
\sin(\alpha^* + \sigma_i \alpha_i) < 0,
\end{aligned}
\label{CoNd__2}
\end{equation}
and we must show that there exists $\hat{m} \in \mathds{Z}$ such that \eqref{Proposition_4} with $\hat{m}$ substituted into \eqref{alpha_star_soln} is equivalent to \eqref{CoNd__2}. Choosing $\hat{m} = m+2M-n$ yields the desired result, and therefore statement (b) of the proposition is established. The characterization of the associated equilibrium values follows from the preceding discussion\editKG{, and also establishes the sufficiency of our existence conditions, completing the proof}.
\end{proof}
\begin{remark}
The possibility of having $2M-n= 0$ cannot be ruled out for an even number of agents, in which case \eqref{cond_on_alpha_star} can only be satisfied if $\sum\alpha_i$ is an integer multiple of $\pi$. This case corresponds to existence of a continuum of circling equilibria.
\end{remark}

\begin{remark}
Letting $\psi_{i,i+1}$ denote the angular separation between agent $i$ and $i+1$ at a circling equilibrium, we have
\begin{equation}
\cos\psi_{i,i+1}
=
\frac{\rho_{ib}^2 + \rho_{i+1,b}^2 - \rho_i^2}{2 \rho_{ib} \rho_{i+1,b}}
=\cos(2\kappa_i).
\end{equation}
Therefore the equilibrium value of angular separation between agent $i$ and agent $i+1$ is $2\kappa_i$.
\end{remark}

\begin{remark}
\label{remark:leftbranch}
If we consider the special case for which \eqref{Solution_normal} holds true for each pair of agents (i.e. the leftmost branch in Fig~\ref{fig:tree_1}), then we have $\sigma = (1,1,\dots,1)$, i.e. $M=n$. In this case \eqref{cond_on_alpha_star} simplifies to the form
\begin{equation}
\alpha^* = m \left( \frac{\pi}{n} \right) - \sum_{i=1}^{n} \left( \frac{\alpha_i}{n} \right).
\label{alpha_soln_special_case}
\end{equation}
\end{remark}

\section{Stability analysis for the two-agent system}
\label{sec:2_Agent}
Here we consider the case where $n=2$ and analyze the stability of the associated circling equilibria. For the two-agent system, our constraints \eqref{FINAL_Constraint_1}-\eqref{FINAL_Constraint_2} imply that $\rho_1 = \rho_2 \triangleq \rho$ and $\theta_i = \kappa_i$ for $i=1,2$. Thus the dynamics \eqref{CL_dynamics_n_simplified} simplify for the two-agent case to 
\begin{align}
\label{CL_dynamics_2_simplified}
\dot{\rho}
&= 
- (\cos\kappa_1 + \cos\kappa_2)
\nonumber \\
\dot{\kappa}_i 
&= 
- \mu \Big[(1 - \lambda)\sin(\kappa_i - \alpha_i) + \lambda \sin(\kappa_{ib} - \alpha_0) \Big] 
+ \frac{\lambda}{\rho}(\sin\kappa_1 + \sin\kappa_2) \\
\dot{\rho}_{ib} 
&= 
- \cos\kappa_{ib}
\nonumber \\
\dot{\kappa}_{ib} 
&= 
\dot{\kappa}_i
- \frac{1}{\rho}(\sin\kappa_1 + \sin\kappa_2)
+ \frac{1}{\rho_{ib}} \sin\kappa_{ib},  \; i=1,2,
\nonumber
\end{align}
and are subject to the \editKG{constraint \eqref{FINAL_Constraint_1} which simplifies to} 
\begin{align}
\rho \mathds{I}_2 - \rho_{1b} R(\kappa_{1b} - \kappa_1) - \rho_{2b} R(\kappa_{2b} - \kappa_{2}) &= 0. 
\label{2_agent_CoNsTrAiNt_simplified}
\end{align}

\subsection{Existence of circling equilibria for the two-agent case}

For the two-agent system, we have only two possible branches in Fig~\ref{fig:tree_1}. The right-hand branch corresponds to $M=1$, for which we have $2M-n = 0$, and \textit{Proposition \ref{prop:existenceProp}} does not apply. However, if $\alpha_1 + \alpha_2 = \editKG{k}\pi$ for some $\editKG{k} \in \mathds{Z}$, then a continuum of equilibria exist with $\kappa_1$ and $\kappa_2$ satisfying $\kappa_1 + \kappa_2 = \pi + \alpha_1 + \alpha_2$. 

For the left-hand branch in Fig~\ref{fig:tree_1}, we have $M=2$ (i.e. $2M-n \neq 0$) and therefore we can apply \textit{Proposition \ref{prop:existenceProp}} (as long as $\sin(\sum \alpha_i) \neq 0$) for which \eqref{alpha_star_soln} simplifies to 
\begin{equation}
\alpha^* = m \left( \frac{\pi}{2} \right) - \sum_{i=1}^{2} \left( \frac{\alpha_i}{2} \right).
\label{alpha_soln_special_case_n2}
\end{equation}
\editKG{Thus we have (for $i=1,2$)
\begin{equation}
\sin(\alpha^* + \sigma_i \alpha_i) 
= 
\sin\left(m\frac{\pi}{2} + \alpha_i - \frac{\alpha_1+\alpha_2}{2}\right),
\end{equation}
and the second constraint} in \eqref{Proposition_4} requires $m= \pm 1$. We label these options as Type 1 and Type 2 two-agent circling equilibrium and summarize the resulting characterization in Table~\ref{Table_1_FULL}, where we have made use of the notation
\begin{align}
	\alpha^+ = (\alpha_1+\alpha_2)/2, \quad \alpha^- = (\alpha_1-\alpha_2)/2.
\end{align}

\subsection{Stability analysis}
We analyze the stability of two-agent circling equilibria by linearizing the dynamics \eqref{CL_dynamics_2_simplified} about the equilibria described in the previous section. Following the line of thought from the stability analysis in \cite{Marshall_TAC_04}, we first demonstrate that the linearized dynamics will always have exactly one pair of pure imaginary eigenvalues resulting from the constraint equation \eqref{2_agent_CoNsTrAiNt_simplified}. 

First, denoting $\xi = \left\{\kappa_1, \kappa_2,\rho,  \kappa_{1b}, \kappa_{2b},\rho_{1b},  \rho_{2b}\right\}$ and the corresponding dynamics \eqref{CL_dynamics_2_simplified} by $f(\xi)$, we express \eqref{2_agent_CoNsTrAiNt_simplified} in terms of scalar constraints by
\begin{align}
	g_1(\xi) &\triangleq \rho - \rho_{1b}\cos(\kappa_{1b}-\kappa_1) - \rho_{2b}\cos(\kappa_{2b}-\kappa_2) = 0, \nonumber \\
	g_2(\xi) &\triangleq \rho_{1b}\sin(\kappa_{1b}-\kappa_1) + \rho_{2b}\sin(\kappa_{2b}-\kappa_2) = 0. \nonumber
\end{align}
We define the manifold on which these constraints are satisfied by 
\begin{align}
	M = \left\{\xi \in \mathbb{R}^{7n}: g_1(\xi) = g_2(\xi) = 0\right\},
\end{align}
which can be shown to be invariant under the dynamics \eqref{CL_dynamics_2_simplified}. If we let $\bar{\xi}$ denote a representative circling equilibrium for the two-agent case and let $\dot{\tilde{\xi}} = A \tilde{\xi}$ denote the linearization of the dynamics \eqref{2_agent_CoNsTrAiNt_simplified} about $\bar{\xi}$, then it was demonstrated in \cite{Marshall_TAC_04} that invariance of $M$ implies existence of a change of basis which will transform $A$ into upper-triangular form with a $2\times 2$ lower-right hand block. A suitable explicit form for the change of basis is given by $\phi = \Phi(\xi)$, where 
\begin{align}
	\phi_1 = \kappa_1, \phi_2=\kappa_2, \phi_3=\rho, \phi_4=\kappa_{1b}, \nonumber \\
	\phi_5=\rho_{1b},\phi_6=g_1(\xi),\phi_7=g_2(\xi).
\end{align}
We note that the corresponding equilibrium $\bar{\phi} = \Phi(\bar{\xi})$ will have $0$'s for the last two components.

By a straightforward calculation, we have
\begin{align}
\label{eqn:dDotEqns}
	\dot{g}_{1}(\xi) &= \frac{\partial g_1(\xi)}{\partial \xi}f(\xi) = -\frac{\sin \kappa_1 + \sin \kappa_2}{\rho}g_2(\xi), \nonumber \\
	\dot{g}_{2}(\xi) &= \frac{\partial g_2(\xi)}{\partial \xi}f(\xi) = \frac{\sin \kappa_1 + \sin \kappa_2}{\rho}g_1(\xi), 
\end{align}
from whence it follows that, under the change of basis, the dynamics linearized about $\bar{\phi}$ take the upper-triangular form
\begin{align}
	\dot{\phi} = \left[\begin{array}{cc} 
	  A_{11} & * \\
		0_{2\times 5} & A_{22}
		\end{array}\right]\phi,
\end{align}
with 
\begin{align}
\label{eqn:A22matrix}
	A_{22} &= \left[\begin{array}{cc} 
	  0 & -\frac{\sin \kappa_1 + \sin \kappa_2}{\rho} \\
		\frac{\sin \kappa_1 + \sin \kappa_2}{\rho} & 0	\end{array}\right]_{\xi = \Phi^{-1}(\phi)} \nonumber \\
		&= \left[\begin{array}{cc} 
	  0 & \mp \delta \\
		\pm \delta & 0 \end{array}\right],		
\end{align}
where $\delta \triangleq \mu\left(\cos\alpha_0 + \left(\frac{1-\lambda}{\lambda}\right)\cos\alpha^{+}\right)$. It is clear from \eqref{eqn:A22matrix} that $A_{22}$ has a pair of pure imaginary eigenvalues at $\lambda = \pm j\mu\delta$ (resulting from the constraint equation \eqref{2_agent_CoNsTrAiNt_simplified}). By analogy with the argument presented in \cite{Marshall_TAC_04}, we focus our stability characterization on the remaining five eigenvalues.  

Returning to the original coordinates $\xi$, we proceed with our stability analysis by linearizing the dynamics about the Type 1 CCW circling equilibrium from Table \ref{Table_1_FULL}. One can show that the corresponding characteristic polynomial $P(x)$ is given by
\begin{align}
	P(x) &= (x^2 + \delta^2)(x^2 +\lambda\editKG{\Psi} x + \lambda \delta^2)
	\bigl(x^3 + \lambda\Psi x^2 + \delta^2 x + (1-\lambda)\mu \sin\alpha^{+} \delta^2\bigr),
\end{align}
where 
\begin{align}
	\Psi \triangleq \mu\left[\sin(\alpha_0) + \left(\frac{1-\lambda}{\lambda}\right)\sin(\alpha^{+})\right].
\end{align}
As expected based on the previous discussion, $P(x)$ has two pure imaginary roots at $x = \pm j\delta$. It is also clear that the roots of the quadratic term will have strictly negative real part if and only if $\Psi>0$. By the Routh-Hurwitz criterion, the roots of the cubic factor will be in the open left-half plane if and only if $\Psi>0$, $\sin\alpha^{+}>0$, and $\lambda\Psi\delta^2 - (1-\lambda)\mu \sin\alpha^{+} \delta^2 > 0$. Since this last condition simplifies to $\lambda\Psi - (1-\lambda)\mu \sin\alpha^{+} > 0$, which always holds if $\sin(\alpha_0) > 0$, our requirement for stability is given by $\sin\alpha^{+}>0$ and $\sin(\alpha_0) > 0$. 

A similar analysis of the other possible circling equilibria (i.e. Type 1 CW and Type 2 CCW and CW) results in analogous stability conditions, which can be summarized in the following proposition.

\begin{proposition} 
\label{prop:TwoAgentStability}
The Jacobian associated with the two-agent circling equilibria has two pure imaginary eigenvalues resulting from the constraint equation \eqref{2_agent_CoNsTrAiNt_simplified}. The remaining eigenvalues all have real parts less than zero if and only if 
\begin{itemize}
	\item $\sin(\alpha_0) > 0$ and $\sin \alpha^{+}>0$ in the Type 1 CCW case;
	\item $\sin(\alpha_0) < 0$ and $\sin \alpha^{+}<0$ in the Type 1 CW case;	
	\item $\sin(\alpha_0) > 0$ and $\sin \alpha^{+}<0$ in the Type 2 CCW case; 
	\item $\sin(\alpha_0) < 0$ and $\sin \alpha^{+}>0$ in the Type 2 CW case.
\end{itemize}
\end{proposition}

\vspace {.25cm}

\begin{proof}
Follows from the discussion above.
\end{proof}

Note that these results are also summarized in Table~\ref{Table_1_FULL}.

%
%
\begin{table*}[t!]
\centering
\begin{adjustbox}{max width=\columnwidth}
\vspace{-0.1in}
\begin{tabular}{lllll}
\hline
& 
\multicolumn{2}{c}{Type 1} 
& 
\multicolumn{2}{c}{Type 2}
\\
\hline
$\begin{array}{l}
\textrm{\textcolor{my_GREEN}{Existence Condition}}
\end{array}$
&
\multicolumn{2}{c}{
$\begin{array}{l}
\cos\alpha^- > 0
\\
\lambda \cos \alpha_0 + (1 - \lambda) \cos\alpha^+ > 0
\end{array}$
} 
& 
\multicolumn{2}{c}{
$\begin{array}{l}
\cos\alpha^- < 0
\\
\lambda \cos \alpha_0 - (1 - \lambda) \cos\alpha^+ > 0
\end{array}$
}
\\
\hline
& 
$CCW$ 
& 
$CW$ 
& 
$CW$ 
& 
$CCW$
\\
\cmidrule(l){2-5}
$\begin{array}{l}
\textrm{\textcolor{blue}{Chracterization}}
\\
\textrm{\textcolor{blue}{of}}
\\
\textrm{\textcolor{blue}{Equilibria}}
\end{array}$
& 
$\begin{array}{l}
\kappa_{ib} = \frac{\pi}{2}
\\
\kappa_1 = \frac{\pi}{2} + \alpha^-
\\
\kappa_2 = \frac{\pi}{2} - \alpha^-
\\
\psi_{12} = \pi + 2\alpha^-
\end{array}$
&
$\begin{array}{l}
\kappa_{ib} = -\frac{\pi}{2}
\\
\kappa_1 = -\frac{\pi}{2} + \alpha^-
\\
\kappa_2 = -\frac{\pi}{2} - \alpha^-
\\
\psi_{12} = - \pi + 2\alpha^-
\end{array}$
&
$\begin{array}{l}
\kappa_{ib} = -\frac{\pi}{2}
\\
\kappa_1 = \frac{\pi}{2} + \alpha^-
\\
\kappa_2 = \frac{\pi}{2} - \alpha^-
\\
\psi_{12} = \pi + 2\alpha^-
\end{array}$
&
$\begin{array}{l}
\kappa_{ib} = \frac{\pi}{2}
\\
\kappa_1 = -\frac{\pi}{2} + \alpha^-
\\
\kappa_2 = -\frac{\pi}{2} - \alpha^-
\\
\psi_{12} = -\pi + 2\alpha^-
\end{array}$
\\
\cmidrule(l){2-5}
&
\multicolumn{2}{l}{
$\begin{array}{l}
\displaystyle \rho_i = \frac{2\lambda \cos\alpha^-}{\mu\lambda \cos \alpha_0 + \mu(1 - \lambda) \cos\alpha^+}
\\
\displaystyle \rho_{ib} = \frac{\lambda}{\mu \lambda \cos \alpha_0 + \mu (1 - \lambda) \cos\alpha^+}
\end{array}$
} 
&
\multicolumn{2}{l}{
$\begin{array}{l}
\displaystyle  \rho_i = \frac{2\lambda \cos\alpha^-}{\mu(1 - \lambda) \cos\alpha^+ - \mu \lambda \cos \alpha_0}
\\
\displaystyle \rho_{ib} = \frac{\lambda}{\mu (1 - \lambda) \cos\alpha^+ - \mu \lambda \cos \alpha_0 }
\end{array}$
}
\\
\hline
$\begin{array}{l}
\textrm{\textcolor{red}{Stability Condition}}
\end{array}$
&
$\begin{array}{l}
\sin \alpha_0 > 0
\\
\sin \alpha^+ > 0
\end{array}$
&
$\begin{array}{l}
\sin \alpha_0 < 0
\\
\sin \alpha^+ < 0
\end{array}$
&
$\begin{array}{l}
\sin \alpha_0 > 0
\\
\sin \alpha^+ < 0
\end{array}$
&
$\begin{array}{l}
\sin \alpha_0 < 0
\\
\sin \alpha^+ > 0
\end{array}$
\\
\hline
\end{tabular}
\end{adjustbox}
\caption{Characterization of circling equilibria for a two-agent system (whenever $\sin(2\alpha^+) \neq 0$), with $\alpha^+ \triangleq (\alpha_1 + \alpha_2)/2$ and $\alpha^- \triangleq (\alpha_1 - \alpha_2)/2$.}
\label{Table_1_FULL}
\end{table*}
%
%
\section{Implementation Results}
\label{sec:Implementation}
\subsection{Experimental setup}
We use Pioneer 3 DX (from Adept MobileRobots), a compact differential-drive mobile robot with reversible DC motors, high-resolution motion encoders, as the experimental platform. Onboard computation is done via 32-bit Renesas SH2-7144 RISC microprocessor, including the P3-SH microcontroller with ARCOS. \editKG{ARIA, a software library from the developer, provides an interface for controlling and receiving data from the robot, and communication with the robot for sending control commands (\textit{forward velocity} and \textit{turning rate}) is carried out via 802.11-b/g/n networking.}
\begin{figure}[h]
\begin{center}
  \includegraphics[width=0.35\textwidth]{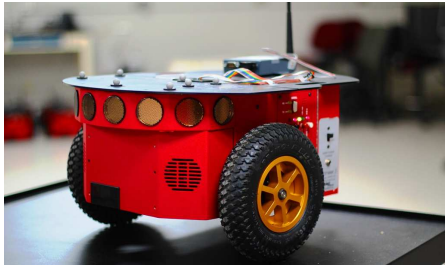}
  \caption{Mobile robot based experimental platform (Pioneer 3 DX) with two-wheel differential and caster.} 
  \label{Robot_P3_DX}
\end{center}
\end{figure}

Algorithm implementation (i.e, feedback law computation) has been \editKG{carried out} in C++ using ROS (Robot Operating System), along with ROS-ARIA, as the interfacing robotics middleware. The experiments have been carried out in a laboratory environment equipped with a sub-millimeter accurate Vicon motion capture system (\textit{www.vicon.com}). The Dell workstation, which we use to evaluate control commands at $25Hz$, is connected to the Vicon server via a dedicated Ethernet connection. 
\subsection{Two robots with asymmetric distribution on the circle}
\begin{figure}[!ht]
        \centering
        \begin{subfigure}[b]{0.5\textwidth}
                \centering
                \includegraphics[width=0.9\textwidth]{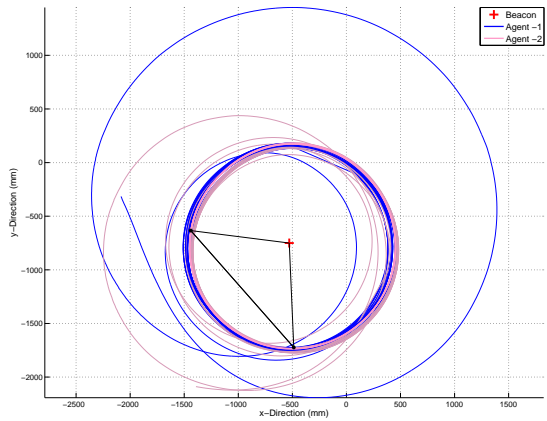}
                \caption{Robot trajectories}
                \label{fig:2_1}
        \end{subfigure}
 
        \begin{subfigure}[b]{0.5\textwidth}
                \centering
                \includegraphics[width=0.9\textwidth]{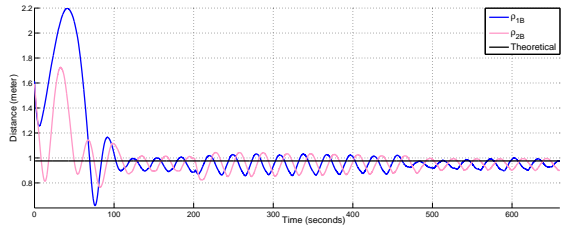}
                \caption{Distances from the beacon}
                \label{fig:2_2}
        \end{subfigure}
        
        \begin{subfigure}[b]{0.5\textwidth}
                \centering
                \includegraphics[width=0.9\textwidth]{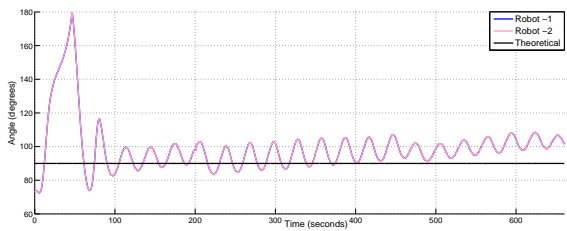}
                \caption{Inter-agent angular separations}
                \label{fig:2_3}
        \end{subfigure}
        \caption{Robot trajectories during implementation of \eqref{CL_dynamics_n_simplified}, along with evolution of relevant quantities (Note that, $n=2$, $\alpha_0 = \pi/3$, $\alpha_1 = 5\pi/12$, $\alpha_2 = -\pi/12$, $\lambda = 1/2$, $\mu = 0.75 m^{-1}$).}\label{fig:2_Agent}
\end{figure}

The first experiment presented here involves two robots circling around the beacon in a counter-clockwise direction. The parameters $\alpha_1$ and $\alpha_2$ were selected as $5\pi/12$ and $-\pi/12$, respectively. These choices, by yielding the equilibrium value of angular separation as $\pi/2$, demonstrate that the proposed approach can give rise to asymmetric distribution of agents on the circle (at equilibrium). Moreover, the parameter values $\alpha_0 = \pi/3$ and $\lambda = 1/2$ yield the circling radius as
\begin{displaymath}
\rho_{ib} = \frac{1}{\mu\big( \cos(\pi/3) + \cos(\pi/6) \big)}, \quad i=1,2, 
\end{displaymath}
and by choosing $\mu = 0.75 m^{-1}$ we get $\rho_{ib} = 0.9761 m$. The corresponding robot trajectories are shown in Fig~\ref{fig:2_1}, and the evolution of distance and angular separation for the agents are shown in Fig~\ref{fig:2_2} and Fig~\ref{fig:2_3}, respectively (refer \cite{ISL_Videos} for implementation videos). These figures show a quick convergence to the circling equilibrium (within $150 sec.$).
%
%
\begin{figure}[!hb]
        \centering
        \begin{subfigure}[b]{0.5\textwidth}
        		\centering
                \includegraphics[width=0.9\textwidth]{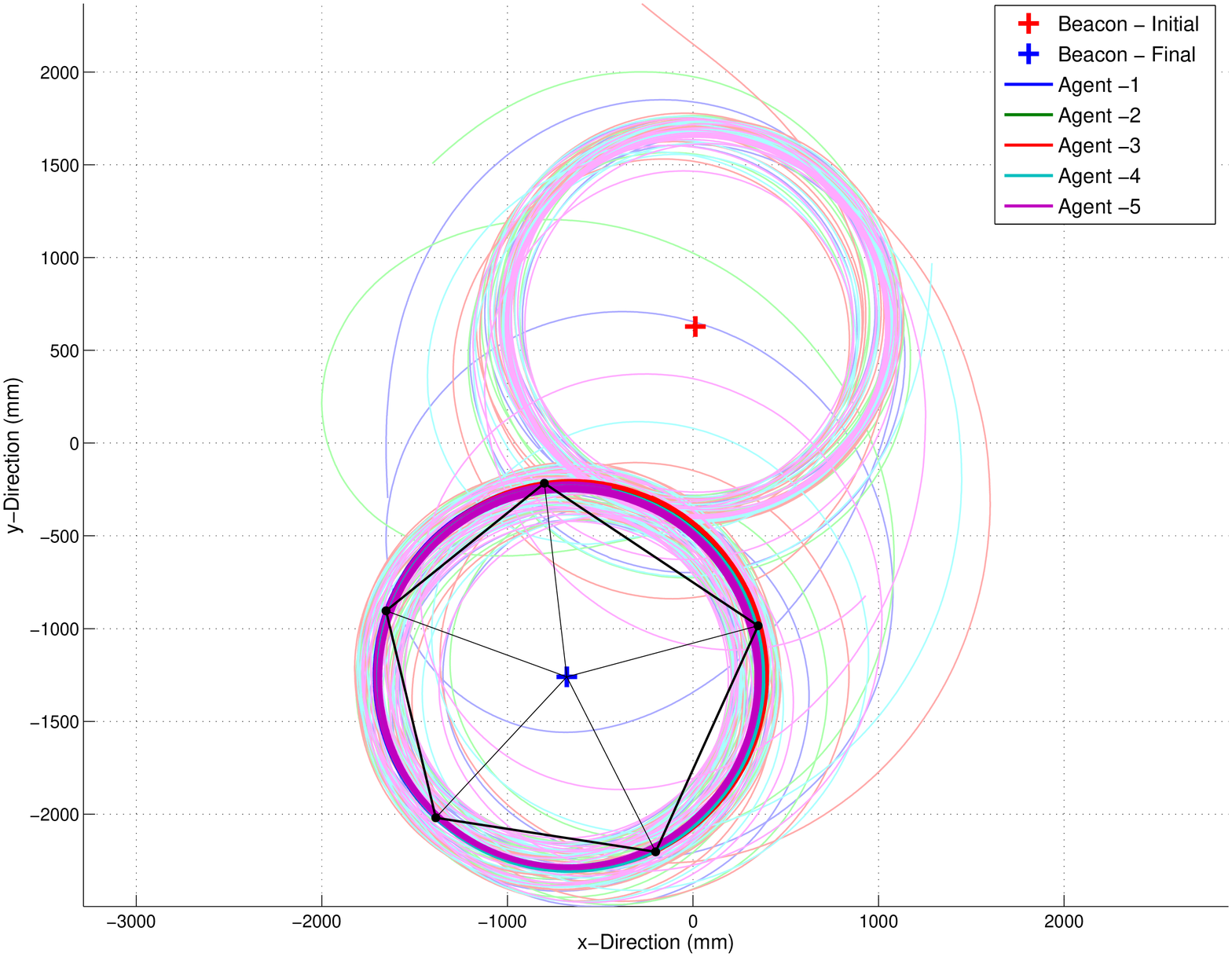}
                \caption{Robot trajectories}
                \label{fig:5_1}
        \end{subfigure}
         
        \begin{subfigure}[b]{0.5\textwidth}
                \centering
                \includegraphics[width=0.9\textwidth]{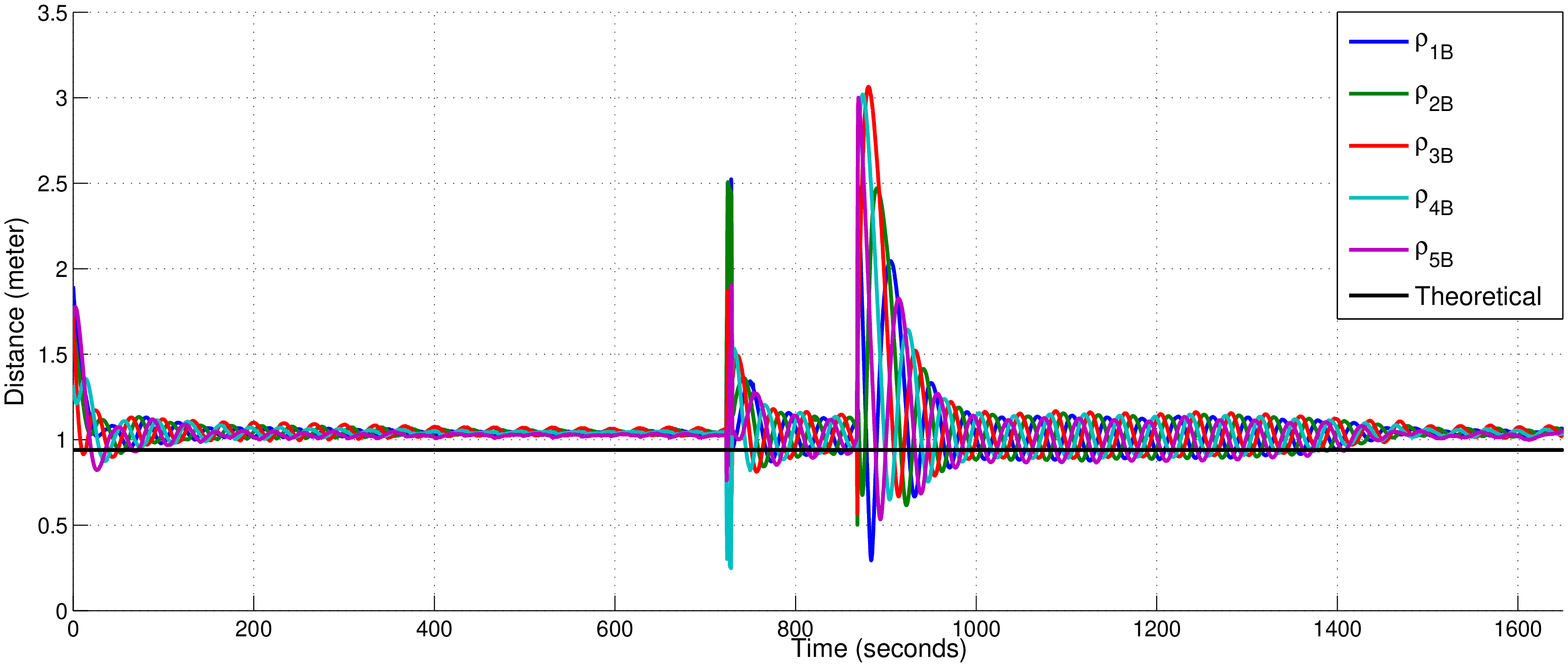}
                \caption{Distances from the beacon}
                \label{fig:5_2}
        \end{subfigure}
         
        \begin{subfigure}[b]{0.5\textwidth}
                \centering
                \includegraphics[width=0.9\textwidth]{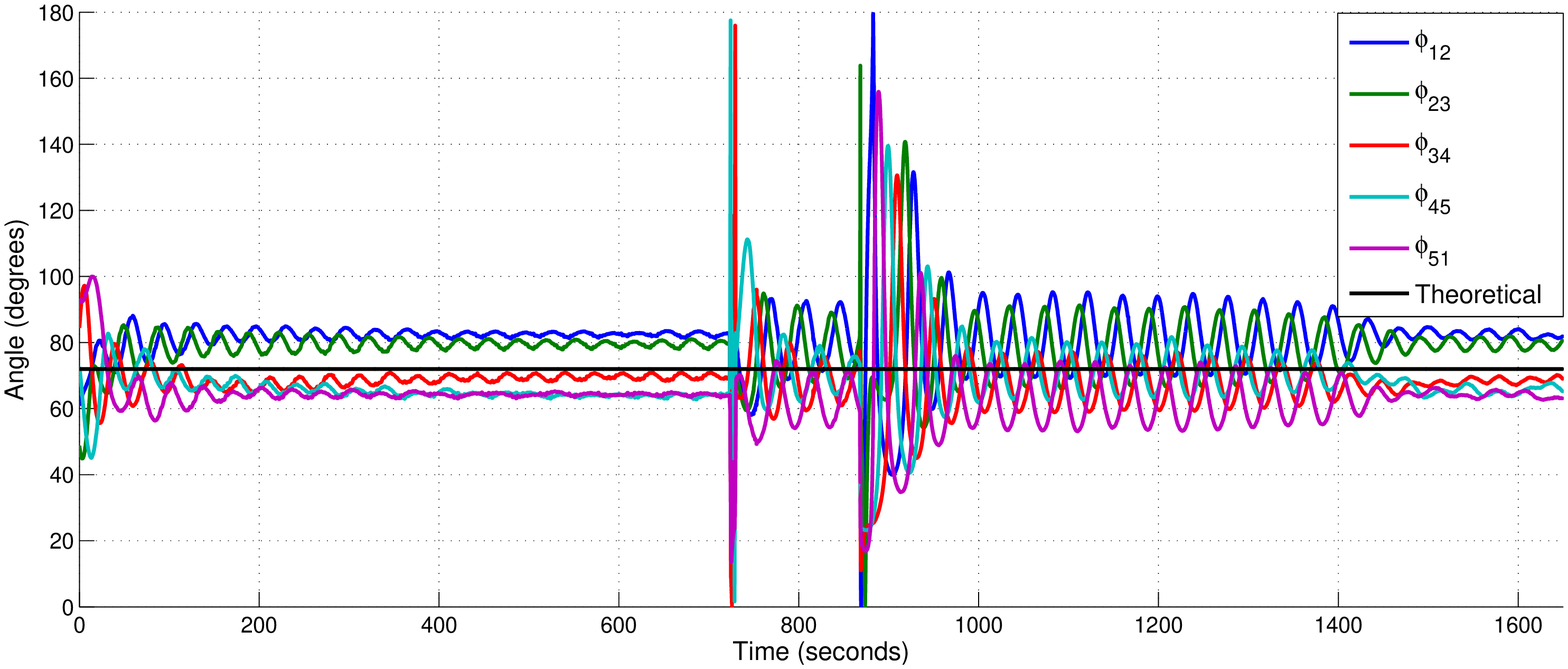}
                \caption{Inter-agent angular separations}
                \label{fig:5_3}
        \end{subfigure}
        \caption{Robot trajectories during implementation of \eqref{CL_dynamics_n_simplified}, along with evolution of relevant quantities (Note that, $n=5$, $\alpha_0 = -\pi/6$, $\alpha_i = -\pi/4$, $\lambda = 1/2$, $\mu = 1.50 m^{-1}$). \editKG{A perturbation was applied to the system at $723 sec.$ into the experiment, and at $868 sec.$ the beacon was relocated to a new position.}}\label{fig:5_Agent}
\end{figure}
%

%
%
\subsection{Five robots with symmetric distribution on the circle}
Next we \editKG{choose control parameters which result in} five robots circling around the beacon in a clockwise direction\editKG{, with the robots distributed symmetrically around the circle. More specifically, we choose the same value for every $\alpha_i$ ($=-\pi/4$), and let $\alpha_0 = -\pi/6$ and $\lambda = 1/2$. Then} we have
\begin{displaymath}
\rho_{ib} = \frac{1}{\mu\big( \cos(-\pi/6) - \sin(-\pi/5 + \pi/4) \big)}, \quad i=1,\ldots,5,
\end{displaymath}
and a choice of $\mu = 1.50 m^{-1}$ yields \editKG{an equilibrium circle radius of} $\rho_{ib} = 0.9395 m$. The corresponding robot trajectories, along with the evolution of distance and angular separation of the agents, are shown in Fig~\ref{fig:5_Agent} (refer \cite{ISL_Videos} for implementation videos). In this experiment we introduced a perturbation to the system at $723 sec.$ into the experiment, and later (at $868 sec.$) the beacon has been moved to a new position. In both cases, the formation quickly converges back to the desired circling equilibrium.
\subsection{Discussion}
The results show some level of imperfection during implementation of the proposed feedback law. This deviation from theoretical predictions can be attributed to multiple factors. To start with, our theoretical analysis assumes the agents to be point particles where in reality they occupy significant space (width - $380 mm$, swing radius - $260 mm$). Also the placement of markers (necessary for measurement using Vicon motion capture system) introduces some error due to misalignment between center of the robot axle and origin of the body fixed frame. Furthermore, as only planar components of positions and heading were considered in measuring relevant quantities, a small slope (which has later been verified) in the lab floor acts as another source of error. However, in spite of these multiple sources of error, the proposed feedback mechanism \eqref{u_i_shape} is able to restrict the error margins well  below the physical dimension of the agents.
%
%
%
%
\section{Conclusion and Future Work}
\label{sec:Concl}
We have introduced a modified version of the CB pursuit law which references a fixed beacon as well as a neighboring agent, and demonstrated that implementation in a cycle graph (with ``spokes'') yields an interesting set of closed-loop dynamics. Analysis of those dynamics reveals the existence of circling equilibria centered on the beacon, and it is of particular interest that a specific equilibrium radius emerges as a function of the control parameters. Future work will focus on extending our analysis to the 3-d setting, as well as consideration of scenarios with multiple beacons or slowly moving beacons. 
%
%
\section*{Acknowledgments}
The authors would like to take this opportunity to thank P. S. Krishnaprasad and E. W. Justh for their valuable feedback and comments. They also appreciate the assistance from U. Halder in implementing the control law on mobile robots.

%
\bibliographystyle{IEEEtran}
\bibliography{ACC_refs}

\begin{thebibliography}{10}
\providecommand{\url}[1]{#1}
\csname url@rmstyle\endcsname
\providecommand{\newblock}{\relax}
\providecommand{\bibinfo}[2]{#2}
\providecommand\BIBentrySTDinterwordspacing{\spaceskip=0pt\relax}
\providecommand\BIBentryALTinterwordstretchfactor{4}
\providecommand\BIBentryALTinterwordspacing{\spaceskip=\fontdimen2\font plus
\BIBentryALTinterwordstretchfactor\fontdimen3\font minus
  \fontdimen4\font\relax}
\providecommand\BIBforeignlanguage[2]{{%
\expandafter\ifx\csname l@#1\endcsname\relax
\typeout{** WARNING: IEEEtran.bst: No hyphenation pattern has been}%
\typeout{** loaded for the language `#1'. Using the pattern for}%
\typeout{** the default language instead.}%
\else
\language=\csname l@#1\endcsname
\fi
#2}}

\bibitem{Kevin_2011_CDC}
K.~S. Galloway, E.~Justh, and P.~S. Krishnaprasad, ``Portraits of cyclic
  pursuit,'' in \emph{Proceedings of 50th {IEEE} {C}onference on {D}ecision and
  {C}ontrol and {E}uropean {C}ontrol {C}onference ({CDC-ECC})}, Orlando,
  Florida, 2011, pp. 2724--2731.

\bibitem{Galloway_PRS_13}
K.~S. Galloway, E.~W. Justh, and P.~S. Krishnaprasad, ``Symmetry and reduction
  in collectives: cyclic pursuit strategies,'' \emph{Proceedings of the Royal
  Society A: Mathematical, Physical and Engineering Science}, vol. 469, no.
  2158, 2013.

\bibitem{Marshall_TAC_04}
J.~A. Marshall, M.~E. Broucke, and B.~A. Francis, ``Formations of vehicles in
  cyclic pursuit,'' \emph{IEEE Transactions on Automatic Control}, vol.~49,
  no.~11, pp. 246--251, 2004.

\bibitem{Kim20071426}
T.-H. Kim and T.~Sugie, ``{Cooperative Control for Target-Capturing Task Based
  on a Cyclic Pursuit Strategy},'' \emph{Automatica}, vol.~43, no.~8, pp. 1426
  -- 1431, 2007.

\bibitem{Smith20051045}
S.~L. Smith, M.~E. Broucke, and B.~A. Francis, ``{A Hierarchical Cyclic Pursuit
  Scheme for Vehicle Networks},'' \emph{Automatica}, vol.~41, no.~6, pp. 1045
  -- 1053, 2005.

\bibitem{Marshall20063}
J.~A. Marshall, M.~E. Broucke, and B.~A. Francis, ``{Pursuit Formations of
  Unicycles},'' \emph{Automatica}, vol.~42, no.~1, pp. 3 -- 12, 2006.

\bibitem{5160735}
J.~Ramirez, M.~Pavone, E.~Frazzoli, and D.~Miller, ``{Distributed Control of
  Spacecraft Formation via Cyclic Pursuit: Theory and Experiments},'' in
  \emph{Proceedings of the {A}merican {C}ontrol {C}onference ({ACC})}, June
  2009, pp. 4811--4817.

\bibitem{Wei_Justh_PSK_09}
E.~Wei, E.~W. Justh, and P.~S. Krishnaprasad, ``Pursuit and an evolutionary
  game,'' \emph{Proceedings of the Royal Society A: Mathematical, Physical and
  Engineering Science}, vol. 465, no. 2105, pp. 1539--1559, 2009.

\bibitem{Seeley_Behav_Ecol_91}
T.~D. Seeley, S.~Camazine, and J.~Sneyd, ``Collective decision-making in honey
  bees: how colonies choose among nectar sources,'' \emph{Behavioral Ecology
  and Sociobiology}, vol.~28, no.~4, pp. 277--290, 1991.

\bibitem{Justh03steeringlaws}
E.~Justh and P.~S. Krishnaprasad, ``Steering laws and continuum models for
  planar formations,'' in \emph{Proceedings of 42nd {IEEE} {C}onference on
  {D}ecision and {C}ontrol ({CDC})}, Maui, Hawaii, 2003, pp. 3609--3615.

\bibitem{Justh_PSK_SCL04}
E.~W. Justh and P.~S. Krishnaprasad, ``Equilibria and steering laws for planar
  formations,'' \emph{Systems \& Control Letters}, vol.~52, no.~1, pp. 25 --
  38, 2004.

\bibitem{Nat_Frenet_Bishop}
R.~L. Bishop, ``There is more than one way to frame a curve,'' \emph{The
  American Mathematical Monthly}, vol.~82, no.~3, pp. 246--251, 1975.

\bibitem{ISL_Videos}
\BIBentryALTinterwordspacing
{Implementation Videos}. [Online]. Available: \url{http://ter.ps/beaconcb}
\BIBentrySTDinterwordspacing

\end{thebibliography}

\end{document}